\documentclass[journal,12pt,onecolumn]{IEEEtran}
\usepackage{cite}
\usepackage{url}
\usepackage[cmex10]{amsmath}
\usepackage{amssymb,amsthm}

\newcommand\bF{\mathbb F}
\newcommand\bP{\mathbb P}

\newcommand{\bbsm}{\left (\begin{smallmatrix}}      \newcommand{\besm}{\end{smallmatrix}\right )}

\newtheorem{theorem}{Theorem}[section]

\newtheorem{proposition}[theorem]{Proposition}

\newtheorem{definition}[theorem]{Definition}
\newenvironment{pf}[1][Proof]{\begin{trivlist}
\item[\hskip \labelsep {\bfseries #1}]}{\end{trivlist}}

\newenvironment{remark}[1][Remarks]{\begin{trivlist}
\item[\hskip \labelsep {\bfseries #1}]}{\end{trivlist}}

\begin{document}
\title{Weight spectrum of codes associated with the  Grassmannian $G(3,7)$}

\author{Krishna~Kaipa,
        Harish~K.~Pillai%, ~\IEEEmembership{Member,~IEEE,}
        % <-this % stops a space
\thanks{K. Kaipa is with the department of Mathematics, and H.K. Pillai is with the department of Electrical Engineering at the Indian Institute of Technology, Bombay.}} 

%\thanks{Manuscript received January 19, 2012; revised February 11, 2012.}}
%\markboth{IEEE Transactions on Information Theory}
%{Shell \MakeLowercase{\textit{et al.}}: Weight Spectrum of $G(3,7)$ Grassmann codes}

\maketitle

\begin{abstract}
In this paper we consider the problem of determining the weight spectrum of $q$-ary codes $C(3,m)$ associated with Grassmann varieties $G(3,m)$.
For $m=6$ this was done in \cite{Nogin2}. We derive a formula for the weight of a codeword  of $C(3,m)$, in terms of certain varieties associated with alternating trilinear forms on $\bF_q^m$. The classification of such forms under the action of the general linear group $GL(m,\bF_q)$ is the other component that is required to calculate the spectrum of $C(3,m)$. For $m=7$, we explicitly determine the varieties mentioned above. The classification problem for alternating $3$-forms on $\bF^7$ was solved in \cite{Cohen} which we then use to determine the spectrum of $C(3,7)$.
\end{abstract}

\IEEEpeerreviewmaketitle

\section{Introduction}

\IEEEPARstart{G}{rassmann} codes are linear codes associated with the Grassmann variety $G(\ell,m)$ of $\ell$ dimensional subspaces of an $m$ dimensional vector space $\bF^m$, where $\bF$ is a finite field with $q$ elements. They were first studied by Ryan \cite{Ryan} for $q=2$, and Nogin \cite{Nogin} for general $q$. These codes are conveniently described using the correspondence between non-degenerate $[n,k]_q$ linear codes on one hand and non-degenerate $[n,k]$ projective systems on the other hand \cite{Tsfasman}. A non-degenerate $[n,k]$ projective system is simply a collection of $n$ points in projective space $\bP^{k-1}$ satisfying the condition that no hyperplane of
$\bP^{k-1}$ contains all the $n$ points under consideration. \\

The projective system used to define the Grassman codes $C(\ell,m)$ is given by the classical Pl\"{u}cker embedding of $G(\ell,m)$ in $\bP (\wedge^{\ell} \bF^m)$, the projective $\ell$-th exterior power of $\bF^m$. Thus the parameters $n$ and $k$ of the codes $C(\ell,m)$ are :
%equal to $|G(l,m)|$ and  $\binom{m}{l}$ respectively.
\begin{IEEEeqnarray}{rCl}
n = |G(\ell,m)| &=&\frac{(q^m-1) (q^{m-1}-1) \ldots (q^{m-\ell+1}-1)}{(q^{\ell}-1)(q^{\ell-1}-1) \ldots (q-1)} \nonumber \\
 k &=& \binom{m}{\ell}
\end{IEEEeqnarray}

We briefly recall the Pl\"{u}cker embedding. To an $\ell$ dimensional subspace $\Lambda$ of $\bF^m$ we assign the wedge product  $v_1 \wedge v_2 \wedge \dotsb \wedge v_{\ell}$ where $\{ v_1, \dotsc, v_{\ell}\}$ is an arbitrary basis of $\Lambda$. The expression $v_1 \wedge \dotsb \wedge v_{\ell}$ considered as an element of $\bP (\wedge^{\ell} \bF^m)$ is independent of the choice of basis. This defines a one-one map of $G(\ell,m)$ in $\bP (\wedge^{\ell} \bF^m)$. The image of this map is a non-singular variety defined by the Pl\"{u}cker relations.
% and is an embedding of $G(l,m)$ into $\bP(\wedge^l \bF^m)$.
Let $\{e_1, \dotsc, e_m\}$ be a basis of $\bF^m$ and let $\mathcal I(\ell,m)$ denote the set of multi-indices:
\[ \mathcal I(\ell,m) = \{ (i_1,\cdots,i_{\ell}) \,|\, 1 \leq i_1 < i_2 < \dotsb < i_{\ell} \leq m\}\]
 Then $\{ e_I \, | \, I \in  \mathcal I(\ell,m) \}$ is a basis of $\wedge^{\ell} \bF^m$. In terms of this basis, the Pl\"{u}cker image of the $\ell$ dimensional subspace $\Lambda$ is given as
\[ \sum_{I \in  \mathcal I(\ell,m)}  p_I  \, e_I \]
where the $k$ homogeneous coordinates $p_I$ are called the Pl\"{u}cker coordinates of $\Lambda$.
A hyperplane $\mathcal H$ in  $\bP (\wedge^{\ell} \bF^m)$ is given by a linear  equation:
$\sum_{I \in \mathcal I(\ell,m)} c_I p_I = 0$.
It is clear that if $c_I \neq 0$, then the $\ell$-plane having basis $\{e_i \,|\, i \in I \}$ does not lie in $\mathcal H$. This shows that the Pl\"{u}cker embedding is non-degenerate. \\

Assigning some order to the $n$ points $\{P_1, \dotsc, P_n\}$ of $G(\ell,m)$, and also to the $k$ elements of $\mathcal I(\ell,m)$, we form the $k \times n$ matrix $M$ whose entries $M_{ij}$ are given by the
$i$-th  Pl\"{u}cker coordinate of $P_j$. The matrix $M$ is the generator matrix of the code $C(\ell,m)$.
The non-degeneracy condition implies that the matrix has full rank. If we left-multiply $M$ by a message-word (a row vector $(a_1, \dotsc, a_k)$ of length $k$), we obtain a codeword (a row vector $(b_1, \dotsc, b_n)$ of length $n$) in a one-one manner. Thus $M$ generates a linear code $\bF^k \hookrightarrow \bF^n$. Observe that the row-span of $M$ is the space of codewords. The entries $b_i$ of the codeword are equal to the values at $P_i$ of the functional on $\wedge^{\ell} \bF^m$, given by $\sum_{I \in  \mathcal I(\ell,m)}  a_I e^I$  where $\{e^I = e^{i_1} \wedge
\dotsb \wedge e^{i_{\ell}}   \,|\, I \in \mathcal I(\ell,m)\}$ is the dual basis to $ \{ e_I \, | \, I \in  \mathcal I(\ell,m) \}$. Thus we see that
%Since the rows of $M$ correspond to coordinate hyperplanes $p_I = 0$, we see that
the message-words correspond to  elements of $\wedge^{\ell} (\bF^m)^*$, the space of functionals on $\wedge^{\ell} (\bF^m)$. There is a bijective correspondence between hyperplanes $\mathcal H$ in $\bP(\wedge^{\ell} \bF^m)$ and points $\omega_{\mathcal H}$ of the projective space of non-zero message-words $\bP(\wedge^{\ell} (\bF^m)^*)$. In this correspondence, the kernel of the functional $\omega_{\mathcal H}$ is precisely $\mathcal H$. We may also think of elements of
$\bP(\wedge^{\ell} (\bF^m)^*)$ as the projective space of alternating $\ell$-multilinear functions (or $\ell$-forms) on $\bF^m$.

%Thus we identify the codewords with %elements of $\bP(\wedge^l (\bF^m)^*)$. \\
 %In this correspondence the zero set of the multilinear function $\omega_{\mathcal H}$ is precisely $\mathcal H$. Thus we identify the codewords with %elements of $\bP(\wedge^l (\bF^m)^*)$. \\

The weight of a codeword corresponding to $\omega_{\mathcal H}$ (i.e its Hamming norm) is simply the number of points $P_i, 1 \leq i \leq n$ not lying on $\mathcal H$. By abuse of notation, we often refer to $\omega_{\mathcal H}$ as the codeword. Consider the function $\omega \mapsto $ wt$(\omega)$ from non-zero codewords to positive integers. The image of this function, together with the number of pre-images for each integer in the image, is called the spectrum of the code $C(\ell,m)$.
%As we vary the codewords ($l$-forms) the various weights that we obtain, together with the number of codewords having that weight,  %are known as the spectrum of the code $C(l,m)$.
The weight of a non-zero codeword $\omega$ only depends on its projective class. Therefore, for determining the spectrum of $C(\ell,m)$, it suffices to consider only the projective space of codewords. \\

The weight spectrum of codes $C(2,m)$ for all $m$, and the weight spectrum of $C(3,6)$ were determined by Nogin in \cite{Nogin} and \cite{Nogin2} respectively.   The organization of the article is as follows. In section \ref{sec:weight_formula} we introduce what we call the weight varieties associated to a $3$-form, and derive a formula for the  weight of a codeword of $C(3,m)$ in terms of the cardinalities of these varieties. The calculation of the spectrum of $C(3,m)$ requires us to determine the possible values of these weights, as well as the number of codewords having each of these weights.
This weight classification of codewords is facilitated by the classification of projective $3$-forms on $\bF^m$ under the action of the projective linear group $PGL(m,\bF)$.
In section \ref{sec:projective_classification}, we obtain this classification for $m=7$, by a minor modification of the results of the authors of \cite{Cohen}.   In section \ref{sec:weight_classification} we determine the weight varieties (and their cardinalities), of representative codewords of this classification, and calculate the spectrum of $C(3,7)$.

\section{A formula for the weight of a $3$-form} \label{sec:weight_formula}
We derive a formula for the weight of a codeword of the code $C(3,m)$.
The following notation will be used in this section.
$V$ denotes the vector space $\bF^m$. For any set $A \subset V \setminus \{0\}$ which satisfies $c \cdot A \subset A$ for all non-zero scalars $c$,  we use the notation $\bP A$ to denote the projectivization of $A$.
For a finite set $S$,  $|S|$ denotes its cardinality.  The cardinality of the general linear group $GL(m,\bF)$ will be denoted by $[m]_q$:
\[   [m]_q = q^{m(m-1)/2} \,  (q^m-1)(q^{m-1}-1) \cdots (q-1) \]

Given a codeword of $C(3,m)$, let $\omega$ be the corresponding $3$-form on $\bF^m$, and let $\mathcal H_{\omega}$  be the corresponding hyperplane of $\bP(\wedge^3 \bF^m)$ as described above. The weight of the codeword $\omega$ is
\[  {\rm wt}(\omega) = \left| \{P_i \,:\, 1 \leq i \leq n, P_i \notin \mathcal H_{\omega} \} \right|  \]
%= n - |G(l,m) \cap \mathcal H_{\omega}|
We will frequently use the following observation:  the total number of ordered bases of all $3$-dimensional subspaces of $V$ represented by the $P_i \notin \mathcal H_{\omega}$ put together,  is  $ [3]_q \, {\rm wt} (\omega) $.  \\

\begin{definition} \label{G-action}
The group $GL(m,\bF)$ acts on $3$-forms by taking a $3$-form $\omega$ to the $3$-form $g \cdot \omega$ defined by
\[ (g \cdot \omega)(v_1,v_2,v_3) = \omega(gv_1,gv_2,gv_3)\]
 For a $3$-form $\omega$ on $\bF^m$ we define {\em Aut}$(\omega)$ to be the group:
%\[ \mathrm{Aut}(\omega) = \{g \in GL(m,\bF) \,|\, g \cdot \omega = c_g \,  \omega \;\; \mathrm{ for \; some } \; c_g \neq 0 \}\]
\[ \mathrm{Aut}(\omega) = \{g \in GL(m,\bF) \,|\, g \cdot \omega =  \omega \}\]
\end{definition}

\subsection{Weight of a degenerate $3$-form}   \label{sec:degenerate}
We consider the map $ \phi_{\omega} : V \to \wedge^{2}  V^*$ sending $ v \mapsto \iota_v \omega$ where $\iota_v$ is the operation of interior multiplication defined by:
\[ \langle \iota_v \omega, \beta \rangle  = \langle \omega, v \wedge \beta \rangle, \quad  \forall \, \beta \in \wedge^2 V  \]
Here $\langle\,,\rangle$ is the pairing between $ \wedge^j V^*$ and $\wedge^j V$ for each $j$.

\begin{definition} We say that the  $3$-form $\omega$ is non-degenerate if {\em ker}$(\phi_{\omega}) = \{0\}$.
\end{definition}
If $\omega$ is degenerate, let  ker$(\phi_{\omega})$ be $r$-dimensional. We pick a basis $\{e_1,\cdots e_m\}$ of $V$ such that that
$\{e_{m-r+1}, \cdots, e_m\}$ is a basis for ker$(\phi_{\omega})$. Let $W$ denote the span of $\{e_1, \cdots, e_{m-r}\}$.
Let $\tilde \omega$  denote the restriction of the form $\omega$ to $W$. Since $W \cap $ ker$(\phi_{\omega}) = \{0\}$, it is clear that
$\tilde \omega$ is a non-degenerate $3$-form on $W$. Thus $\tilde \omega$ can be thought of as a codeword in $C(3,m-r)$.
 %$x_1, \ldots, x_r$ be a basis for {\em Ker}$(\phi_{\omega})$ and let $W$ be a complementary $m-r$ dimensional subspace. Writing $V %= W \oplus $ {\em Ker}$ (\phi_{\omega})$, let $\tilde \omega$ denote the restriction of $\omega$ to $W$.

\begin{proposition}
\label{prop:degenerate_weight}
${\rm wt}(\omega) = q^{3r} \, {\rm wt} (\tilde \omega) \quad$ {\em and}  $\quad |${\em Aut}$(\omega)| = |${\em Aut}$(\tilde \omega)| \, [r]_q \, q^{r(m-r)}$
\end{proposition}

\begin{pf}
%We pick a basis $\{e_1, e_2, \dotsc, e_m\}$ of $V$ such that $e_{m-r+i} = x_i$ and $\{e_1, \dotsc, e_{m-r}\}$ is a basis for $W$.
We have:
\begin{equation} \label{eq:wt_def}
[3]_q \!\cdot {\rm wt}(\omega)  = \left| \{ [v_1, v_2, v_3] \, : \, \langle \omega, v_1 \wedge v_2 \wedge v_3 \rangle \neq 0 \} \right|
\end{equation}
where $[v_1, v_2, v_3]$ denotes a $m \times 3$ matrix with columns $v_1, v_2$, and $v_3$.
Since the expression for $\omega$ in terms of the dual basis $\{e^1,\dotsc,e^m\}$ is independent of $e^{m-r+1}, \ldots, e^m$, the last $r$ rows of the matrix
$[v_1, v_2, v_3]$ are arbitrary. Moreover if $[u_1, u_2, u_3]$ is the submatrix formed by the first $m-r$ rows then :
\begin{IEEEeqnarray*}{rCl}
 [3]_q \!\cdot {\rm wt}(\omega) &=& q^{3r} \,\left|\{ [u_1, u_2, u_3] \, : \, \langle \tilde\omega, u_1 \wedge u_2 \wedge u_3 \rangle \neq 0 \}\right| \\
 &=&  q^{3r} \cdot  [3]_q \, {\rm wt} (\tilde \omega)
\end{IEEEeqnarray*}

If $g \in $ Aut$(\omega)$ then the equation $ \omega(g v_1,g v_2,gv_3) = \omega(v_1,v_2,v_3)$ implies that $\iota_{g v_1} \omega$  is zero if and only if $\iota_{v_1} \omega = 0$. Thus Aut$(\omega)$ carries ker$(\phi_{\omega})$ to itself.  Therefore,
with respect to the basis $\{e_1, \cdots,e_m\}$, we can write $g = \bbsm \tilde g & 0 \\ k & h \besm$.
Such a matrix $g$ is in Aut$(\omega)$ if and only if $\tilde g \in$ Aut$(\tilde \omega)$, $h \in GL(r,\bF)$, and $k$ an arbitrary $(m-r) \times r$ matrix. Thus the cardinality of Aut$(\omega)$ is $[r]_q \, q^{(m-r)r}$ times the cardinality of Aut$(\tilde \omega)$.
%(As a  group Aut$(\omega)$ is the semidirect product $($Aut$(\tilde \omega) \times GL(r,\bF) ) \rtimes $ Mat$(m-r \times r, \bF)$)
 \qed
\end{pf}
%The
The proposition shows that in order to calculate the weights of codewords of $C(3,m)$, it is enough to know only the weights of non-degenerate codewords of $C(3,\tilde m)$ for $\tilde m \le m$. The cardinality of Aut$(\omega)$ is useful in determining the number of codewords having a given weight (i.e the spectrum). 
%can be determined in terms of the cardinality of Aut$(\omega)$ (as we demonstrate later for the case $m=7$).
%(For $m=6$ this is how the spectrum was calculated in \cite{Nogin2}.)
%Thus the  above proposition reduces the problem of determining the spectrum of $C(3,7)$ to that of $C(3,\tilde m)$ for $\tilde m <m$  %and to the problem of determining Aut$(\omega)$ for non-degenerate codewords.
%Of course, knowing $C(3,m-1)$ is equivalent to knowing the spectrum of $C(3,m-r)$.

\subsection{Weight varieties of a non-degenerate $3$-form}
Let $V$ be an $m$-dimensional vector space over an arbitrary field $F$. Given a $2$-form $\lambda \in \wedge^2 V^*$, we define certain quantities Pf$_k(\lambda) \in \wedge^{2k} V^*$, for each $k \geq 1$  which we call the  $k$-th Pfaffian of $\lambda$. Let Pf$_0(\lambda) = 1$.
We define Pf$_k(\lambda) \in \wedge^{2k} V^*$ inductively by requiring:
\begin{equation} \label{eq:pfaff_defn}
\iota_v \lambda \wedge {\rm Pf}_{k-1}(\lambda)  = \iota_v {\rm Pf}_k(\lambda), \quad \forall \, v \in V
\end{equation}
This Pf$_k(\lambda)$ generalizes the forms $ \tfrac{\lambda^k}{k!} = \tfrac{1}{k!} (\lambda \wedge \cdots \wedge \lambda
)$, which are used over the fields of real and complex numbers,  to fields with arbitrary characteristic.
We recall the following standard  diagonalization theorem (see \cite{Lang}, section $XV.8$) for $2$-forms on $V$.
The rank of a $2$-form $\lambda$  is the rank of the matrix whose $(i,j)$-th entry is $\lambda(e_i,e_j)$  for any basis $\{e_1, \cdots,e_m\}$ of $V$. The rank is an even integer $2 r$, and one can always pick a basis of $V$ such that the associated matrix is block diagonal with $r$ blocks consisting of  the $2 \times 2$ matrix $ \bbsm 0 & 1\\ -1 &0 \besm$ and zeros elsewhere.

\begin{proposition} [$k$-th Pfaffians of a $2$-form]
\label{prop:pfaffians} \hfill \begin{enumerate}
\item Given a  $\lambda \in \wedge^2 V^*$, $\lambda \neq 0$, for each $k \geq 1$  there is a unique element {\em Pf}$_k(\lambda) \in \wedge^{2k} V^*$
satisfying \eqref{eq:pfaff_defn}.
\item \begin{equation} \label{eq:pfaff_add}
 {\rm Pf}_k(\lambda_1+\lambda_2)= \sum_{j=0}^k {\rm Pf}_j(\lambda_1) \wedge {\rm Pf}_{k-j}(\lambda_2) \end{equation}
\item The unique integer $2r$ such that {\em Pf}$_r(\lambda) \neq 0$ and  {\em Pf}$_{r+1}(\lambda) = 0$, is the rank of $\lambda$.
\end{enumerate}
\end{proposition}

\begin{pf} 
We assume inductively that the  first assertion of the proposition holds for  $1 \leq j \leq k-1$.
The uniqueness of Pf$_k(\lambda)$ follows from the fact that any form $\alpha$ is determined completely by the forms $\{ \iota_v \alpha \,|\, v \in V\}$.  As for existence, we consider the function
\[ f(v_1, v_2, \dotsc, v_{2k}) = \langle \iota_{v_1} \lambda \wedge {\rm Pf}_{k-1}(\lambda), v_2 \wedge \dotsb \wedge v_{2k} \rangle \]
The function $f$ is clearly multilinear. It is also alternating in the variables $v_2, \dotsc, v_{2k}$. In order to prove that
$f$ is a $2k$-form on $V$, it suffices to show that $f(v_1, v_1, v_3, \dotsc, v_{2k}) = 0$ :
\begin{IEEEeqnarray*}{rCl}
f(v_1, v_1, v_3, &\dotsc&, v_{2k}) = \langle \iota_{v_1} \lambda \wedge {\rm Pf}_{k-1}(\lambda), v_1 \wedge v_3 \dotsb \wedge v_{2k} \rangle \\
 &=& \langle - \iota_{v_1} \lambda \wedge \iota_{v_1}{\rm Pf}_{k-1}(\lambda), v_3 \wedge \dotsb \wedge v_{2k} \rangle \\
 &=& \langle - \iota_{v_1} \lambda \wedge \iota_{v_1} \lambda \wedge {\rm Pf}_{k-2}(\lambda), v_3 \wedge \dotsb \wedge v_{2k} \rangle  \\
 &=& 0 \quad \; (\mbox{because }\;   \iota_{v_1} \lambda \wedge \iota_{v_1} \lambda = 0)
\end{IEEEeqnarray*}
It also follows from the definition of $f$ that  $\iota_{v_1}f =  \iota_{v_1} \lambda \wedge {\rm Pf}_{k-1}(\lambda)$ thus proving that Pf$_k(\lambda) = f$. \\

The second assertion easily follows from the defining equation \eqref{eq:pfaff_defn}  and induction. To prove the third assertion, we observe that if  Pf$_j(\lambda) = 0$, then by \eqref{eq:pfaff_defn},  Pf$_i(\lambda) = 0$ for all $i \geq j$. Since Pf$_1(\lambda) = \lambda \neq 0$, there is a unique integer $2\rho$ satisfying Pf$_\rho(\lambda) \neq 0$ and  Pf$_{\rho+1}(\lambda) = 0$.

Using a special basis $\{e_1, \cdots, e_n\}$ of $F^m$ as in the diagonalization theorem mentioned before the proposition, we write
\[ \lambda = e^1 \wedge e^2 + e^3 \wedge e^4 +\dotsb+ e^{2r-1} \wedge e^{2r}\] where $2r$ is the rank of $\lambda$ and $\{e^1, \cdots, e^m\}$ is the dual basis. Using this expansion of $\lambda$ in \eqref{eq:pfaff_add}, we calculate all the $k$-th Pfaffians of $\lambda$, and  find that $\rho = r$.
\end{pf}

\begin{remark} If char$(F)$ does not divide $k!$, then by uniqueness Pf$_k(\lambda)$ is simply $\tfrac{\lambda^k}{k!}$.
The equations Pf$_2(\lambda) = 0$ are the Pl\"{u}cker equations defining decomposable elements of $\bP(\wedge^2 V^*)$, or in other words the Pl\"{u}cker embedding of $G(2, V^*)$ in $\bP (\wedge^2 V^*)$. Given a skew symmetric $2k \times 2k$ matrix $A$ with entries in $F$,
(with diagonal terms required to be zero if char$(F)=2$), 
we can associate a $2$-form to it by $\lambda = \sum_{i<j} A_{ij} e^i \wedge e^j$ where $\{e^1,\cdots, e^{2k}\}$ is the dual basis to the standard basis of $F^{2k}$. Then Pf$_k(\lambda)$ equals $e^1\wedge\cdots\wedge e^{2k}$ times a scalar  Pf$(A)$  (whose square equals det$(A)$) known as the Pfaffian of the matrix $A$ (\cite{Lang}, section $XV.9$). We also mention the fact Pf$_k(g \cdot \lambda) = g \cdot $Pf$_k(\lambda)$ for all $g \in GL(m, F)$, which can easily be proved by induction starting from the case $k=1$ and the defining equation \eqref{eq:pfaff_defn}.

\end{remark}

\begin{definition} Given a non-degenerate $3$-form $\omega$ on $\bF^m$, the $k$-th weight variety of $\omega$ is the subvariety of $\bP^{m-1}$ given by:
\[ X_k(\omega) = \bP \{x \in \bF^m\setminus\{0\}  \,|\,  {\rm Pf}_{k+1}(\iota_{x} \omega) = 0 \}\]
We have \[\varnothing = X_0(\omega) \subset X_1(\omega) \subset \dotsb \subset X_{ \lfloor \frac{m-1}{2} \rfloor}(\omega)  = \bP^{m-1}\]
\end{definition}

We will need Nogin's result on spectrum of $C(2,m)$:
\begin{theorem}[Nogin \cite{Nogin}] \label{nogin's formula}
The weight of a codeword in $C(2,m)$ depends only on the rank of its associated $2$-form $\omega$. If rank$(\omega)$ is $2r$, where
$1 \leq r \leq \lfloor \frac{m}{2} \rfloor$, then:
\[ {\rm wt} (\omega) = q^{2(m-r-1)} \, \frac{q^{2r} - 1}{q^2 - 1} \]
\end{theorem}
For each of these $\lfloor \tfrac{m}{2} \rfloor$ weights, the number of codewords of $C(2,m)$ of that weight
is also determined in [2]. We do not need it here.

\begin{theorem}
\label{weight theorem}
Given a non-degenerate $3$-form $\omega$ on $\bF^m$, let \[ n_i : = |X_i(\omega)| - |X_{i-1}(\omega)| \]
The weight {\em wt}($\omega$) is given by:
\begin{equation} \label{weight formula}
{\rm wt}(\omega) = \frac{q^{2m-4}}{(q^2-1)(1+q+q^2)} \sum_{i = 1}^{\lfloor \frac{m-1}{2} \rfloor} n_i (1 - q^{-2i})
\end{equation}
\end{theorem}
\begin{pf} For any $v_1 \neq 0$,  let   $\{e_1, e_2, \dotsc, e_m\}$ be a basis of $V$ such that $e_1 = v_1$. Let $W$ denote the subspace generated by  $\{e_2, \dotsc, e_m\}$, and let $\pi: V \to W$ be the projection on the last $m-1$ coordinates.  Let $\omega_{v_1}$ be the $2$-form on $W$ obtained by restricting $\iota_{v_1} \omega$  to $W$.
A pair of vectors $v_2, v_3 \in V$ satisfy  $ \langle \iota_{v_1} \omega, v_2 \wedge v_3 \rangle   \neq 0$ if and only if $  \langle \omega_{v_1}, \pi(v_2) \wedge \pi(v_3) \rangle   \neq 0$. Since the first components of $v_2, v_3$ are arbitrary, the cardinality of such pairs  $\{v_2, v_3\} $ is:  \[ q^2 \, [2]_q  \, \mbox{wt}(\omega_{v_1})\]
We thus have:
\begin{IEEEeqnarray*}{rCl}
[3]_q \!\cdot {\rm wt}(\omega)  &=& \left| \{ [v_1, v_2, v_3] \, : \, \langle \omega, v_1 \wedge v_2 \wedge v_3 \rangle \neq 0 \} \right| \\
&=&  \sum_{v_1 \neq 0} \left| \{ [v_2, v_3] \, : \,  \langle \iota_{v_1}\omega,  v_2 \wedge v_3 \rangle \neq 0 \} \right| \\
&=&  q^2 \,  [2]_q  \sum_{v_1 \neq 0} {\rm wt}(\omega_{v_1})
\end{IEEEeqnarray*}

In the sum over all $v_1 \neq 0$, there are $(q-1) n_i$  terms for which the line through  $v_1$  is in  $X_i(\omega) \setminus X_{i-1}(\omega)$.  For such a $v_1$,  $\omega_{v_1}$ has rank $2i$ as a $2$-form on $W$.
By Theorem \ref{nogin's formula}
\[ {\rm wt}(\omega_{v_1}) =  q^{2(m-1 - i-1)} \, \frac{q^{2i}-1}{q^2-1} \]
Substituting this expression for wt$(\omega_{v_1})$ above, we get
\begin{IEEEeqnarray*}{rCl}
  {\rm wt}(\omega)&=& \frac{q^2  [2]_q (q-1)}{[3]_q}  \sum_{i = 1}^{\lfloor \frac{m-1}{2} \rfloor} n_i \, \frac{q^{2i}-1}{q^2-1} \,q^{2m - 4 - 2i} \\
&=& \frac{q^{2m-4}}{(q^2-1)(1+q+q^2)} \sum_{i = 1}^{\lfloor \frac{m-1}{2} \rfloor} n_i (1 - q^{-2i})
\end{IEEEeqnarray*} \qed
\end{pf}

For later use, we specialize formula  \eqref{weight formula} to the cases $m=6, 7$.
For the case $m=6$, we use  $n_1+n_2 = | \bP^5|$ in \eqref{weight formula} to get
\begin{equation} \label{eq:wt6}
 {\rm wt}(\omega) =  q^4 \left[( q^5+q^3+q^2+1)  - \frac{n_1}{1+q+q^2} \right]
\end{equation}
For the case $m=7$, we use $n_1+n_2 +n_3= | \bP^6|$  in \eqref{weight formula} to get
\begin{IEEEeqnarray}{rCl}
\label{eq:wt7}
 {\rm wt}(\omega) =  q^4 \left[ ( q^8+q^6+q^5+q^4+q^3+q^2+1)  \right.  \nonumber \\
  -  \left. \frac{n_2+n_1(1+q^2)}{1+q+q^2} \right]
\end{IEEEeqnarray}

\subsection{The variety $X_2(\omega)$  of a non-degenerate $3$-form on $\bF^7$}
%Nogin has already calculated the weight spectrum of $C(3,6)$ \cite{Nogin2}. Therefore, we demonstrate the above for calculating the weight spectrum of $C(3,7)$. 
Let $V = \bF^7$ and $\omega \in \wedge^3 V^*$. We show that the variety $X_2(\omega) \subset \bP^6$ is a quadric hypersurface
given by the vanishing of an explicitly determined  quadratic form $Q_{\omega}$ on $V$. \\
Let $\eta$ be a basis   of the $1$-dimensional space $\wedge^7 V^*$, and let $H_{\eta}: \wedge^6 V^* \to V$ be the isomorphism defined by:
\begin{equation} \label{eq:H_eta}
 \alpha \wedge \beta  = \langle \beta, H_{\eta}(\alpha)  \rangle \, \eta,  \quad \forall \alpha \in \wedge^6 V^*, \beta \in V^* 
\end{equation}
 Let $x \in V \setminus\{0\}$ and let $\alpha = $ Pf$_3(\iota_x \omega)$. We claim that $H_{\eta}(\alpha)$ is a scalar multiple of $x$. Since $\alpha$ is trilinear in $x$, the scalar multiple is a quadratic form $Q(x)$. Pick $y \in \wedge^6 V$ so that $\langle \eta, x \wedge y \rangle = 1$.
For any $\beta \in V^*$ with $\langle \beta, x \rangle = 0$, we will show that $\langle \beta, H_{\eta}(\alpha) \rangle  = 0$, thus proving that $H_{\eta}(\alpha)$ is a scalar multiple of $x$. By the definition of $H_{\eta}$, and choice of $y$ it follows that:
\begin{IEEEeqnarray*}{rCl}
\langle \beta, H_{\eta}(\alpha) \rangle = \langle \langle \beta, H_{\eta}(\alpha) \rangle \eta, x\wedge y \rangle = \langle \alpha \wedge \beta, x \wedge y \rangle  \nonumber \\
 = \langle \iota_x (\alpha \wedge \beta),y \rangle = \langle \iota_x \alpha \wedge \beta, y \rangle
\end{IEEEeqnarray*}

%\[ \langle \beta, H_{\eta}(\alpha) \rangle = \langle \langle \beta, H_{\eta}(\alpha) \rangle \eta, x\wedge y \rangle = \langle \alpha %\wedge \beta, x \wedge y \rangle  =
%\langle \iota_x (\alpha \wedge \beta),y \rangle = \langle \iota_x \alpha \wedge \beta, y \rangle\]
where in the last equality, we have used the fact that $\iota_x \beta = \langle \beta, x \rangle = 0$.
Using the defining property \eqref{eq:pfaff_defn} of the $3$-Pfaffian, and the fact that $\alpha = $Pf$_3(\iota_x \omega)$,  it follows that $\iota_x \alpha = 0$, and hence that
\[ H_{\eta}({\rm Pf}_3(\iota_x \omega)) = Q(x) \cdot x \]

The variety $X_2(\omega) =\bP \{x \in \bF^7\setminus\{0\}  \,|\,  {\rm Pf}_3(\iota_{x} \omega) = 0 \}$ can now be expressed as:
\[X_2(\omega) =\bP \{x \in \bF^7\setminus\{0\}  \,|\,  Q(x)=0\}\]

A different choice $\eta' = a \eta $ for the basis vector of $\wedge^7 V^*$  (where $a$ is a nonzero scalar) gives  an  isomorphism $H_{\eta'} = a^{-1} H_{\eta}$, and hence to the quadratic form $a^{-1} Q(x)$. Since the zero locus of $Q(x)$ and $a^{-1} Q(x)$ is the same, the variety $X_2(\omega)$ does not depend on the choice of $\eta$. We summarize the above discussion:

\begin{theorem} \label{X_2 quadric}
The variety $X_2(\omega) \subset \bP^6$ associated with a non-degenerate $3$-form $\omega$ on $\bF^7$ is a quadric hypersurface given by the vanishing of a quadratic form $Q_{\omega}$ on $\bF^7$.  The  form $Q_{\omega}$ is defined by
\begin{equation} \label{eq:quadratic}
 H_{\eta} ({\rm Pf}_3(\iota_{x} \omega) ) = Q_{\omega}(x) \cdot x  \quad \forall x \in \bF^7  \end{equation}
where $H_{\eta}: \wedge^6 (\bF^7)^* \to \bF^7$ is the linear isomorphism defined in \eqref{eq:H_eta} 
\end{theorem}

{\bf Remark:} Applying $H_{\eta}$ to the equation:
\[ 6 \,{\rm Pf}_3(\iota_x \omega) = (\iota_x \omega)^3 = \iota_x( \omega \wedge \iota_x \omega  \wedge \iota_x \omega)\]
we get $ 6 \,Q_{\omega}(x)\, \eta =\omega \wedge \iota_x \omega  \wedge \iota_x \omega$. If char$(\bF) \neq 2,3$, this relation defines $Q(x)$. If char$(\bF)=2,3$ we have to use \eqref{eq:quadratic} to define $Q(x)$.

%{\bf Remark:} The quadratic form $Q_{\omega}(x)$ satisfies the relation $6 \,Q_{\omega}(x) \, \eta =\omega \wedge \iota_x \omega  %\wedge \iota_x \omega $. If char$(\bF) \neq 2,3$, this relation defines $Q(x)$. If char$(\bF)=2,3$ we have to use \eqref{eq:quadratic} to %define $Q(x)$.

\section{$PGL(7,\bF)$ classification of $3$-forms on $\bF^7$} \label{sec:projective_classification}
%The weight of a codeword of $C(3,7)$ associated with a non-degenerate $3$-form $\omega$ can be computed using \eqref{eq:wt7}, %once the cardinalities $n_1$ and $n_1+n_2$ of the weight varieties are known.  If two nondegenerate $3$-forms have isomorphic weight %varieties then they have the same weights.
Let $G$ denote $GL(7,\bF)$. We recall the action of $G$ on $3$-forms as given in definition ~\ref{G-action}.
\begin{definition} \label{proj_equiv}
We say nonzero $3$-forms $\omega_1$ and $\omega_2$ are projectively equivalent if there is a $g \in G$ and a non-zero scalar $c$ such that $g \cdot \omega_1 = c \, \omega_2$.
We denote this equivalence relation as $\omega_1 \sim \omega_2$.
\end{definition}
Equivalently $\omega_1 \sim \omega_2$, if their projective classes are in the same orbit under the $\bar G = PGL(7,\bF)$ on
$\bP( \wedge^3 (\bF^7)^*)$ induced by the $G$-action on $\wedge^3 V^*$. Here $PGL(7,\bF) = GL(7,\bF)/\bF^*$ is the projective linear group and $\bF^*$ denotes the subgroup in $GL(7,\bF)$ of scalar matrices.

Let $\omega_1 \sim \omega_2$ with $g \cdot \omega_1 = c \omega_2$. Then $\omega_2(v_1,v_2,v_3) = 0$ if and only if $\omega_1(g\cdot v_1,g \cdot v_2,g \cdot v_3)=0$.  The formula \eqref{eq:wt_def} then implies that wt$(\omega_1) = $ wt$(\omega_2)$. Therefore, in order to determine the possible weights of all codewords, it suffices to restrict the classification to
projective equivalence classes of $3$-forms.  There is also a notion of linear equivalence obtained by requiring $c=1$ in Definition \ref{proj_equiv}. The linear equivalence classes of non-zero $3$-forms on $\bF^7$
and their cardinalities were determined by Cohen and Helminck \cite{Cohen}. By grouping together linear classes which have representatives differing by a scalar multiple, we obtain the projective equivalence classes.  The sum of the cardinalities of the linear classes in each such group is equal to $q-1$ times the cardinality of the corresponding projective class. We thus obtain the following theorem:

%It follows from this definition that if $\omega_1 \sim \omega_2$, then  $\iota_{v_1} (c \,\omega_2)$ is equal to the $2$-form defined by %$(g^{-1} v_2, g^{-1}v_3) \mapsto \iota_{g v_1} %\omega_1(v_2, v_3)$.The latter form has the same rank as $\iota_{g v_1} %\omega_1$, and hence $ v_1 \in X_i(\omega_2)$ if and only if $g \cdot v_1 \in X_i(\omega_1) $. Therefore:
%\[ g \cdot  X_i(\omega_2) = X_i(\omega_1) \]
%where $g \cdot X_i (\omega_2)$ refers to the action of $\bar G$ on $\bP V$ induced by the usual action of $G$ on $V$.
\begin{theorem}\label{G-orbits}
% \emph{\cite{Cohen}}
There are eleven projective equivalence classes in $\bP (\wedge^3 (\bF^7)^*)$ with representatives and cardinalities as given below
\begin{IEEEeqnarray}{rCl}
\label{eq:G-orbits}
\nonumber \omega_1&=&  e^{123}\\
 \nonumber \omega_2&=&  e^1 \wedge (e^{23}+e^{45}) \\
 \nonumber \omega_3&=&  e^{123}+e^{456} \\
 \nonumber \omega_4&=&  e^{123}+e^{345} +e^{561} \\
 \nonumber \omega_{5a}&=&  e^1 \wedge (e^{23}+e^{45}) +e^6\wedge (e^{24}+s e^{35}), \,{\rm  if} \, {\rm  char}\, \bF \neq 2\\
 \nonumber  \omega_{5b}&=& e^1 \wedge (e^{23}+e^{45}) +e^6\wedge (e^{24}+s e^{35}+e^{45}), \,{\rm char}\, \bF = 2 \\
\omega_6&=&   e^1 \wedge (e^{23}+e^{45}) +e^{267} \\
 \nonumber \omega_7&=&  e^1 \wedge (e^{23}+e^{45}+e^{67}) \\
 \nonumber \omega_8&=&  e^1 \wedge (e^{23}+e^{57})+e^6 \wedge (e^{27}+e^{45}) \\
  \nonumber \omega_9&=&  e^1 \wedge (e^{23}+e^{45}+e^{67}) +e^{246} \\
 \nonumber \omega_{10}&=&  e^{123}+e^{456} +e^7 \wedge ( e^{14}+e^{25}+e^{36}) \\
 \nonumber \omega_{11a}&=& \omega_{5a} + e^{167}, \,{\rm  if} \, {\rm  char}\, \bF \neq 2\\
 \nonumber \omega_{11b}&=& \omega_{5b} + e^{167},  \,{\rm char}\, \bF = 2
%\\ \nonumber \\
\end{IEEEeqnarray}  \begin{IEEEeqnarray}{rCl}
\nonumber N_1&=&  (q^7-1)(q^5-1)(q^2 -q+1)/ (q-1)^2 \\
\nonumber N_2&=&  q^2(q^7-1)(q^5-1)(q^4+q^2+1)(q^3-1)/ (q-1)^2\\
\nonumber N_3&=&  \tfrac{1}{2} q^9 (q^7-1)(q^5-1)(q^3+1)(q^2+1)/ (q-1)\\
\nonumber N_4&=&  q^4 (q^7-1)(q^6-1)(q^5-1)(q^4-1)/ (q-1)^2\\
\nonumber N_5&=&   \tfrac{1}{2} q^9 (q^7-1)(q^5-1)(q^3-1)(q+1)\\
\nonumber N_6&=&  \tfrac{1}{2} q^9 (q^7-1)(q^6-1)(q^5-1)(q^3-1)(q^2+1)/ (q-1)^2\\
 \label{eq:N_i} N_7&=&  q^6 (q^7-1)(q^5-1)(q^2+q+1)\\
\nonumber N_8&=&  q^{11} (q^7-1)(q^6-1)(q^5-1)(q^2+q+1)(q^2+1)\\
\nonumber N_9&=&  q^6 (q^7-1)(q^6-1)(q^5-1)(q^4-1)/ (q-1)\\
\nonumber N_{10}&=&  q^{15} (q^7-1)(q^5-1)(q^4-1)(q^3-1)\\
\nonumber N_{11}&=&  \tfrac{1}{2} q^9 (q^7-1)(q^6-1)(q^5-1)(q^3-1)
\end{IEEEeqnarray}
\end{theorem}
{\bf Remarks on Theorem \ref{G-orbits}}  The notation $e^{123}$ denotes $e^1 \wedge e^2 \wedge e^3$.
The symbol $s$ above denotes a fixed  element of $\bF$ satisfying the condition that $s$ is not a square if
char $\bF \neq 2$, and that $s$ is not of the form $a(a+1)$ in case char $\bF =2$.
%It is a standard fact about finite fields that any field with $q^2$ elements is isomorphic to
%\begin{equation}
%\bK = \left\{ \,  \begin{IEEEeqnarraybox}[][c]{l?s} \IEEEstrut
%\bF[X]/(X^2-s)  & if char$(\bF) \neq 2$, \\
%\bF[X]/(X^2+s X+1) & if char$(\bF) = 2$
%\IEEEstrut \end{IEEEeqnarraybox} \right.
%\label{eq:K_defn}
%\end{equation}
The number $N_i$  denotes the cardinality  of the projective equivalence class of $\omega_i$.
The linear equivalence class of a $3$-form $\omega$ has cardinality
$|G|/|$Aut$(\omega)|$.  For each $\omega \in \{\omega_1, \cdots, \omega_{11}\}$ except $\omega_{10}$, and any non-zero scalar $c$, the form $c \,\omega$  is linearly equivalent to $\omega$. To see this we just observe that if
$ g \in G$ sends $e_2,e_5,e_7$ to $e_2/c, e_5/c,e_7/c$ and fixes the other basic vectors, then $g \cdot \omega_j = \omega_j$ for $j=1,2,3,4,5,7,9,11$, whereas if
$g \in G$ sends $e_1,e_6$ to $e_1/c, e_6/c$ and fixes the other basic vectors, then $g \cdot \omega_j = \omega_j$ for $j=6,8$.
Thus the cardinalities $N_j$ of the projective classes $\omega_j$, for $j \neq 10$, are obtained by dividing the cardinalities of the linear
equivalence classes by $q-1$. (The forms $\{\omega_1, \omega_2,\omega_3,\omega_4,\omega_5,\omega_6, \omega_7, \omega_8, \omega_9,\omega_{11}\}$ are denoted in \cite{Cohen} by $\{f_1,f_2,f_3,f_4,f_{10}, f_5,f_8,f_7,f_6,f_{11}\}$ respectively.)  The set of all $3$-forms projectively equivalent to $\omega_{10}$ consists of three or one linear equivalence classes according to whether  $3$ divides $q-1$ or not (denoted in tables 1,2 of \cite{Cohen} by $f_9,f_{12,\mu},f_{12,\mu^2}$ in the former case and just $f_9$ in the latter case). However the sum of the cardinalities of these linear equivalence classes is always $(q-1) N_{10}$. The computation of $|$Aut$(\omega_{10})|$  is not elementary. The authors of \cite{Cohen} use the fact that  Aut$(\omega_{10})$ is (possibly upto a cyclic group of order $3$) the automorphism group of the split algebra of Cayley octonions over $\bF$. The latter group is the Chevalley exceptional group $G_2(\bF)$ of order $q^6(q^6-1)(q^2-1)$.  The number $(q-1) N_1$ as calculated in \cite{Cohen} (the first entry of table 2) has a typographical error. The denominator in that expression should be $q^2-1$ instead of $(q-1)^2$.

 Note that $N_1, \cdots,N_5$  can also be calculated using Proposition \ref{prop:degenerate_weight}. If $\omega \in \{\omega_1, \cdots, \omega_5\}$ and $\omega'$ denotes the restriction of $\omega$ to $\bF^6$ (the span of $\{e_1, \cdots, e_6\}$), and $\tilde \omega$ and $r$ are as in the proposition \ref{prop:degenerate_weight}. Then, we have:
\begin{IEEEeqnarray*}{rCl}
|{\rm Aut}(\omega')|&=& |{\rm Aut}(\tilde \omega)| \, [r-1]_q \, q^{(r-1)(m-r)}  \nonumber \\
  |{\rm Aut}(\omega)| &=& |{\rm Aut}(\tilde \omega)| \, [r]_q \, q^{r(m-r)}
\end{IEEEeqnarray*}

%$|$Aut$(\omega')| = |$Aut$(\tilde \omega)| \, [r-1]_q \, q^{(r-1)(m-r)}$ and
%$|$Aut$(\omega)| = |$Aut$(\tilde \omega)| \, [r]_q \, q^{r(m-r)}$.
Therefore  for $1 \leq j \leq 5$, $N_j = 
\tfrac{1}{q-1} |G|/| $Aut$(\omega_j)|$ equals:
 \[ N_j = \frac{ [7]_q\, [r-1]_q}{[6]_q \, [r]_q \,q^{7-r}} \, \frac{A_j}{q-1}\]
where $A_j = |GL(6,\bF)|/|$Aut$(\omega_j)|$ as calculated by Nogin in \cite{Nogin2}.

\section{Weight classification of $3$-forms on $\bF^7$} \label{sec:weight_classification}

% We show in Appendix 1 that  the $PGL(7,\bF)$ orbit with representative $\omega_5$ is independent of the choice of %such an element %$s$. Similarly with $\omega_{11}$.  We first determine the weights of the first five orbits. These %orbits are represented by degenerate %$3$-forms.
%\begin{pf}We first consider the forms $\omega_1 \cdots \omega_5$ as $3$-forms on $\bF^6$. The classes %$\omega_1,\cdots, %\omega_4$ are distinct over the algebraic closure $\bar \bF$ of $\bF$, hence their classes over %$\bF$ are also distinct. The weight of %$\omega_5$ differs from the weights of $\omega_1, \cdots,\omega_4$ as was %shown by Nogin \cite{Nogin2}, therefore it is a new %class. We show below that the class of $\omega_{5a}$

%\begin{proposition} \label{prop: G_3_6_spectrum}
%The weights of $\omega_i$, $1 \leq i \leq 5$  are:
%\begin{IEEEeqnarray*}{rCl}
%{\rm wt}(\omega_1)&=&   q^{12}\\
%{\rm wt}(\omega_2)&=&   q^{12}+q^{10}\\
%{\rm wt}(\omega_3)&=&   q^{12}+q^{10} +  q^9-q^7\\
%{\rm wt}(\omega_4)&=&  q^{12}+q^{10} +  q^9\\
%{\rm wt}(\omega_5)&=&  q^{12}+q^{10} +  q^9+ q^7\\
%\end{IEEEeqnarray*}
%\end{proposition}
%\begin{pf}
%By Proposition~\ref{prop:degenerate_weight},  the weight of $\omega_i$, $1 \leq i \leq 5$  is  $q^3$ times the %weight of $\omega_i$ %viewed as a $3$-form on $\bF^6 = $ span of $\{e_1, \cdots, e_6\}$. The latter weights were %determined in \cite{Nogin2}, from which %our result follows.
%\end{pf}

The weights of the non-degenerate forms $\omega_i$, $ i >5$ can be determined from formula \eqref{eq:wt7} once the cardinalities of the varieties
$X_1(\omega_i)$ and $ X_2(\omega_i)$ are known.  We begin with $X_1(\omega)$. We recall that $X_1(\omega) =
\bP \{ x \in \bF^7  \,|\,  {\rm Pf}_2 (\iota_{x} \omega) = 0 \}$

\begin{proposition} \label{prop:n_1}
The varieties $X_1(\omega_i)$ and their cardinalities  for $i>5$  are:
\begin{IEEEeqnarray*}{rCl}
X_1(\omega_6)&=&\bP^2 \cup_{\bP^0} \bP^2, \quad n_1(\omega_6) = 1+2q+2q^2 \\
X_1(\omega_7)&=&  \bP^5, \quad  n_1(\omega_7) = |\bP^5| \\
%1+q+q^2+q^3+q^4+q^5\\7
X_1(\omega_8)&=&    \bP^1, \quad  n_1(\omega_8) =1+q\\
X_1(\omega_9)&=&   \bP^2, \quad  n_1(\omega_9) =1+q+q^2\\
X_1(\omega_{10})&=& \varnothing, \quad  n_1(\omega_{10}) =0\\
X_1(\omega_{11})&=& \bP^0, \quad  n_1(\omega_{11}) =1
%, \; {\rm if} \, {\rm char}(F) \neq 2 \\ &=& \bP^2, \quad  n_1(\omega_{11}) = |\bP^2|,\;  {\rm if}\, {\rm char} (F)= 2
\end{IEEEeqnarray*}
\end{proposition}

\begin{pf}
Let $x=\sum_{j=1}^{7} x_j e_j$. By Proposition~\ref{prop:pfaffians}, we have
\begin{equation} \label{eq:Pf2eval}
 {\rm Pf}_2(\iota_x \omega) =
\sum_{j=1}^7  x_j^2 \; {\rm Pf}_2(\iota_{e_j} \omega) + \sum_{i<j} x_i x_j \,(\iota_{e_i} \omega) \wedge (\iota_{e_j} \omega)
\end{equation}

We begin with $\omega_6$ and evaluate Pf$_2(\iota_x \omega_6)$ using the above formula \eqref{eq:Pf2eval}. We find that the coefficients of $e^{2345}$ and $e^{3167}$  are $x_1^2$ and $x_2^2$ respectively. Setting these equal to zero we get: \\
\begin{IEEEeqnarray*}{rCl}
\label{eq:n_1(6)}
{\rm Pf}_2(\iota_x \omega_6)_{|x_1=x_2=0} =  e^{12}\wedge(x_4 e^5-x_5 e^4) \wedge(x_7 e^6 -x_6 e^7)
\end{IEEEeqnarray*}
Therefore:
\begin{IEEEeqnarray*}{rCl}
X_1(\omega_6) &=&\{x_1=x_2=0\} \cap \left( \{x_4=x_5=0\} \cup \{x_6=x_7=0\}\right) \\
&=& \bP\{e_3,e_6,e_7\} \cup_{\bP\{e_3\}} \bP\{e_3,e_4,e_5\} \simeq \bP^2 \cup_{\bP^0} \bP^2
\end{IEEEeqnarray*}
Next we consider Pf$_2(\iota_x \omega_7)$. The coefficient of $e^{2345}$ is $x_1^2$, moreover  $x_1$ divides
Pf$_2(\iota_x \omega_7)$. Therefore:
 \[X_1(\omega_7) =\{x_1=0\}  \simeq \bP^5\]
For Pf$_2(\iota_x \omega_8)$, the coefficients of $e^{2357}, e^{1367}, e^{1467}$ and $e^{2457}$ are $x_1^2, -x_2^2, -x_5^2$ and $x_6^2$ respectively. Setting $x_1,x_2,x_5$ and $x_6$ to zero, Pf$_2(\iota_x \omega_8)$ reduces to $(x_3 x_4+x_7^2) e^{1256}$ . Therefore
 \[X_1(\omega_8) =\{x_1=x_2=x_5=x_6= x_3 x_4+x_7^2=0\}\simeq \bP^1\]
The map $(t,s) \mapsto (0,0,t^2,-s^2,0,0,ts)$ establishes an isomorphism between $\bP^1$ and $X_1(\omega_8)$.\\

In Pf$_2(\iota_x \omega_9)$, the coefficients of $e^{2345}, e^{1346}, e^{1526}$ and $e^{1247}$ are $x_1^2, -x_2^2, x_4^2$ and $-x_6^2$ respectively. Setting $x_1,x_2,x_4$ and $x_6$ to zero, Pf$_2(\iota_x \omega_9)$ reduces to $0$. Therefore
 \[X_1(\omega_9) =\{x_1=x_2=x_4=x_6=0\} \simeq \bP^2\]
Next we consider $\iota_x \omega_{10}$ which equals
\begin{IEEEeqnarray*}{rCl}
x_1(e^{47}+e^{23}) +x_2 (e^{57}+e^{31})+x_3(e^{67}+e^{12})
\\ \nonumber +x_4 (e^{71}+e^{56}) +x_5 (e^{72}+e^{64})+x_6 (e^{73}+e^{45}) \\ \nonumber + x_7 (e^{14}+e^{25}+e^{36})
\end{IEEEeqnarray*}

The coefficients of $e^{4723}, e^{5731},e^{6712},e^{7156},e^{7264},e^{7345}$ and $e^{1425}$
in Pf$_2(\iota_x \omega_{10})$  are equal to
$x_1^2,x_2^2,x_3^2,x_4^2,x_5^2,x_6^2$ and $x_7^2$ respectively. Setting these equal to zero we get:
\[ X_1(\omega_{10}) = \varnothing\]

Let $\mu$ be one if char$(\bF)=2$ and zero otherwise. We calculate   $\iota_x \omega_{11}$ to be:
\begin{IEEEeqnarray*}{rCl}
x_1(e^{23}+e^{45}+e^{67}) +x_2 (e^{31}+e^{46})+x_3(e^{12}+s e^{56})
\\ \nonumber+x_4 (e^{51}+e^{62}+\mu e^{56})  +x_5 (e^{14}+s e^{63}+\mu e^{64})\\ \nonumber+x_6 (e^{24}+s e^{35}+\mu e^{45}+e^{71}) + x_7 e^{16}
\end{IEEEeqnarray*}

The coefficients in Pf$_2(\iota_x \omega_{11})$ of $e^{2367}$ and  $e^{2471}$ are $x_1^2$ and $x_6^2$ respectively. Thus $x_1, x_6$ must be zero. The coefficients of  $e^{3146}$ and  $e^{1256}$ are $x_2^2 - s x_5^2 -\mu  \, x_2 x_5 $ and $s x_3^2 - x_4^2 + \mu \,  x_3 x_4$ respectively. By definition of $s \in \bF$, the last two quadratic forms are irreducible and hence, $x_2,x_3,x_4$ and $x_5$ must all be zero for  Pf$_2(\iota_x \omega_{11})$ to vanish. Therefore:
\[X_1(\omega_{11}) = \{x_1 = \cdots =  x_6 = 0\} \simeq   \bP^{0} \] \qed
\end{pf}
We  now compute the varieties $X_2(\omega)$ and their cardinalities.
\begin{proposition} \label{prop:n_2}
The varieties $X_2(\omega_i)$ and their cardinalities  for $i>5$  are:
\begin{IEEEeqnarray*}{rCl}
X_2(\omega_6)&=&\bP^5 \cup_{\bP^4} \bP^5, \quad |X_2(\omega_6)|= 2 |\bP^5| - |\bP^4| \\
X_2(\omega_7)&=&  \bP^5, \quad  |X_2(\omega_7)| = |\bP^5| \\
X_2(\omega_8)&=& (\bP^1 \times \bP^1 \times \bF^3) \amalg \bP^2,\;   |X_2(\omega_8)| =q^3 |\bP^1|^2+|\bP^2|\\
X_2(\omega_9)&=&   \bP^5, \quad  |X_2(\omega_9)| =|\bP^5| \\
X_2(\omega_{10})&=& \overline{SL(\bF^4)/Sp(\bF^4)} , \quad  |X_2(\omega_{10})| = |\bP^5|\\
X_2(\omega_{11})&=& \bP^4, \quad  |X_2(\omega_{11})| =|\bP^4|,
%\; {\rm if} \, {\rm char}(F) \neq 2 \\ &=& \bP^5, %
%\quad  |X_2(\omega_{11})| = |\bP^5|
%,\;  {\rm if}\, {\rm char}(F)= 2
\end{IEEEeqnarray*}
\end{proposition}

\begin{pf}
Let $x=\sum_{j=1}^{7} x_j e_j$. By Proposition~\ref{prop:pfaffians}, we have
\begin{IEEEeqnarray}{rCl} \label{eq:Pf3eval}
 {\rm Pf}_3(\iota_x \omega) =  \sum_{j=1}^7  x_j^3 \, {\rm Pf}_3(\iota_{e_j} \omega)
&+&\\ \nonumber
\sum_{i<j} \, \left[ x_i^2 x_j \,{\rm Pf}_2(\iota_{e_i} \omega) \wedge (\iota_{e_j} \omega)  \right.
&+&  \left. x_i x_j^2 \, (\iota_{e_i} \omega) \wedge{\rm Pf}_2(\iota_{e_j} \omega)\right]
\end{IEEEeqnarray}
Let  $*: \wedge^6 (\bF^7)^* \to \bF^7$ be the  linear isomorphism defined by:
 \[* (e^1 \wedge \cdots \wedge e^{i-1} \wedge e^{i+1} \wedge \cdots \wedge e^7) = (-1)^{i-1} e_i \]
By Proposition ~\ref{X_2 quadric},  there is a unique quadratic form $Q_{\omega}$ on $V$ such that
\[*({\rm Pf}_3(\iota_{x} \omega) ) = Q_{\omega}(x) \cdot x \]
The variety $X_2(\omega)$ is the zero locus of $Q_{\omega}$. To determine $Q_{\omega}$ we expand Pf$_3(\iota_x \omega)$  using
\eqref{eq:Pf3eval}. The quadratic form $Q_{\omega}(x)$ is simply the coefficient of $e^{234567}$ divided by $x_1$.

We calculate Pf$_3(\iota_x \omega_6)$ using the above formula \eqref{eq:Pf3eval} and identify $Q_{\omega_6}(x) = x_1x_2$. Therefore
\begin{IEEEeqnarray*}{rCl}
X_2(\omega_6) &=&\{x_1 x_2=0\} \simeq \bP^5 \cup_{\bP^4} \bP^5
\end{IEEEeqnarray*}
and  $|X_2(\omega_6)|= 2 |\bP^5| - |\bP^4| $. \\

Next we consider Pf$_3(\iota_x \omega_7)$ and calculate $Q_{\omega_7}(x) = x_1^2$. Therefore:
 \[X_2(\omega_7) =X_1(\omega_7) = \{x_1=0\}  \simeq \bP^5\]

Calculating  Pf$_3(\iota_x \omega_8)$, we get $Q_{\omega_8}(x) = x_1 x_5 - x_2 x_6$. The variety $X_2(\omega_8)$ is the disjoint union of the
the subvariety for which at least one  of $x_1,x_2,x_5,x_6$ is non-zero, with the subvariety for which $x_1,x_2,x_5,x_6$ are all zero.
The first subvariety is immediately seen to be $(\bP^1 \times \bP^1) \times \bF^3$ by the Segre embedding, and the second subvariety
is $\bP^2$.
 \[X_2(\omega_8) =\{x_1 x_5 -x_2 x_6= 0\} \simeq (\bP^1 \times \bP^1 \times \bF^3) \amalg \bP^2
\]
and hence  $|X_2(\omega_8)| =q^3 |\bP^1|^2+|\bP^2|$.\\

Next we consider Pf$_3(\iota_x \omega_9)$ and calculate $Q_{\omega_9}(x) = x_1^2$. Therefore:
 \[X_2(\omega_9) = \{x_1=0\}  \simeq \bP^5\]
Calculating  Pf$_3(\iota_x \omega_{10})$, we get $Q_{\omega_{10}}(x) = x_1 x_4 + x_2 x_5 + x_3 x_6   - x_7^2$.\\ The cardinality of
$X_2(\omega_{10})$  can easily be calculated  to be  $|\bP^5|$ (but $X_2(\omega_{10})$ is not $\bP^5$, see below).
We give a description of $X_2(\omega_{10})$ and use it to compute the cardinality.  Writing  $\bP^6 = \bF^6 \amalg \bP^5$ where the affine part $\bF^6$  corresponds to $x_7=1$, and $\bP^5$ is the hyperplane at infinity $x_7=0$, the variety  $X_2(\omega_{10}) \subset \bP^6$ is a disjoint union  $V_0 \amalg V_1$, where  $V_1 = X_2(\omega_{10}) \cap \bF^6$ is the affine part, and $V_0 = X_2(\omega_{10}) \cap \bP^5$ is the part at infinity. $X_2(\omega_{10})$ is thus the projective closure of the affine variety $V_1$.
Comparing with the Pl\"{u}cker relation Pf$_2(\alpha)=0$ defining the Grassmannian $G(2,4)$ of $2$-forms of rank $2$ on $\bF^4$, we see $V_0$ is isomorphic to $G(2,4)$. The variety $V_1$ is isomorphic to the variety:
\[\{ \alpha \in \wedge^2(\bF^4)^* \, |\,  {\rm Pf}_2(\alpha)= e^{1234}\}\]
The formula $g \cdot $ Pf$_2(\alpha) = $ Pf$_2(g \cdot \alpha)$ implies that any $\alpha$ with
${\rm Pf}_2(\alpha)= e^{1234}$ is of the form $g \cdot (e^{12} +e^{34})$ for a $g \in SL(4,\bF)$ uniquely determined upto left multiplication by an element of $Sp(\bF^4)$. Thus $V_1 =  SL(4,\bF)/Sp(\bF^4)$, and  hence $X_2(\omega_{10})$ is the projective closure of the affine variety $SL(\bF^4)/Sp(\bF^4)$).
 \[X_2(\omega_{10}) =\{x_1 x_4 + x_2 x_5 + x_3 x_6 = x_7^2\} \simeq \overline{SL(\bF^4)/Sp(\bF^4)}\]
The cardinality  $|X_2(\omega_{10})|$ equals  $|V_0|+|V_1|$:
 \[  (q^4+q^3+2q^2+q+1) + \frac{q^6 (q^4-1)(q^3-1)(q^2-1)}{q^4(q^4-1)(q^2-1)} = |\bP^5| \]

In the case of  Pf$_3(\iota_x \omega_{11})$  we get $Q_{\omega_{11a}}(x) = x_1^2 - s \, x_6^2$, and
$Q_{\omega_{11b}}(x) = x_1^2 + s \, x_6^2 + x_1 x_6$. By definition of $s \in \bF$ in the cases char$(\bF) \neq 2$, char$(\bF) = 2$  the quadratic forms
$Q_{11a}(x), Q_{11b}(x)$ respectively, are irreducible. Therefore  $Q_{11}(x)$ vanishes  if and only if $x_1$ and $x_6$ both vanish:
\begin{IEEEeqnarray*}{rCl}
X_2(\omega_{11}) &=&\{x_1 =x_6=0\} \simeq \bP^4
%,  \quad \mbox{ if char}(F) \neq 2 \\ \nonumber
%X_2(\omega_{11}) &=&\{x_1 = \sqrt s \, x_6\} \simeq \bP^5,  \quad \mbox{ if char}(F) = 2
\end{IEEEeqnarray*} \qed
\end{pf}

\begin{theorem} \label{wt_theorem}
The weights of $\omega_1, \cdots, \omega_{11}$   are:
\begin{IEEEeqnarray}{rCl}
\label{eq:weights}
\nonumber {\rm wt}(\omega_1)&=&   q^{12}\\ \nonumber
{\rm wt}(\omega_2)&=&   q^{12}+q^{10}\\ \nonumber
{\rm wt}(\omega_3)&=&   q^{12}+q^{10} +  q^9-q^7\\  \nonumber
{\rm wt}(\omega_4)&=&  q^{12}+q^{10} +  q^9\\  \nonumber
{\rm wt}(\omega_{5})&=& q^{12}+q^{10} +  q^9+ q^7\\ \nonumber
{\rm wt}(\omega_6)&=&   q^{12}+q^{10}+q^9+q^8-q^7\\
{\rm wt}(\omega_7)&=&   q^{12}+q^{10}+q^8  \\ \nonumber
{\rm wt}(\omega_8)&=&  q^{12}+q^{10}+q^9+q^8\\ \nonumber
{\rm wt}(\omega_9)&=&  q^{12}+q^{10}+q^9+q^8\\ \nonumber
{\rm wt}(\omega_{10})&=&  q^{12}+q^{10}+q^9+q^8+q^6\\ \nonumber
{\rm wt}(\omega_{11})&=& q^{12}+q^{10}+q^9+q^8+q^7
\end{IEEEeqnarray}
\end{theorem}

\begin{pf}
By Proposition~\ref{prop:degenerate_weight},  the weight of a degenerate form $\omega_i$, $1 \leq i \leq 5$  is  $q^3$ times the weight of $\omega_i$ viewed as a $3$-form on $\bF^6 = $ span of $\{e_1, \cdots, e_6\}$. The latter weights were determined in \cite{Nogin2}.
Multiplying them with $q^3$ we get the weights of $\omega_1,\cdots, \omega_5$. \\
For the nondegenerate forms $\omega_6, \cdots, \omega_{11}$, we use the formula \eqref{eq:wt7} with $n_2(\omega)+n_1(\omega) = |X_2(\omega)|$:
 \begin{IEEEeqnarray*}{rCl}
 {\rm wt}(\omega_i) = q^{12}+q^{10}+q^9+q^8+q^7+q^6+q^4 \nonumber \\
   - \,    q^4 \left(\frac{|X_2(\omega_i)| +q^2 |X_1(\omega_i)| }{1+q+q^2} \right)
\end{IEEEeqnarray*}
The quantities $|X_1(\omega_i)|$ and $|X_2(\omega_i)|$ have been computed in Propositions ~\ref{prop:n_1} and ~\ref{prop:n_2}. Substituting these in the above equation we get the weights of $\omega_6, \cdots, \omega_{11}$. \qed
\end{pf}

We observe that the weights of $\omega_8$ and $\omega_9$  are equal. So we conclude:
\begin{theorem}  The spectrum of the Grassmann code $C(3,7)$ has ten distinct weights: 
\[\{ {\rm wt}(\omega_1), \cdots, {\rm wt}(\omega_8),{\rm wt}(\omega_{10}), {\rm wt}(\omega_{11})\}\]
 where wt$(\omega_i)$ are given in \eqref{eq:weights}.
The number of codewords with weight wt$(\omega_i)$ for $i=1 \cdots8, 10, 11$ are
$q-1$ times $N_1, N_2, \cdots, N_7, N_8+N_9, N_{10}, N_{11}$  respectively, where $N_i$ are  given in  \eqref{eq:N_i}
\end{theorem}
%We remark that the weight varieties and hence the spectrum do not behave differently when $q$ is even.
Let $\mu_1$ denote the number of codewords of the dual code $C(3,7)^{\perp}$ which have weight $1$.
The non-degeneracy of $C(3,7)$ implies that $\mu_1=0$ . The MacWilliams identities can be used to express $\mu_1$ in terms of $N_1, \cdots N_{11}$ and the Krawtchouk polynomial $K_1(x) = (q-1)n-qx$ (\cite{Sloane} p.129,  or \cite{Tsfasman} p.19).
We get:
\[0\,=\, \frac{K_1(0)}{q-1}\,+\,\sum_{i=1}^{11} \,N_i \,K_1({\rm wt}(\omega_i)) \]
Simplifying this equation we get:
\begin{equation} \label{eq:MacWilliams}
\sum_{i=1}^{11} N_i \, {\rm wt}(\omega_i) = q^{34}\,  n  \end{equation}
Using $\sum_{i=1}^{11} N_i = \tfrac{q^{35}-1}{q-1} = |\bP^{34}|$, and $n = |G(3,7)|$, we can rewrite the above equation as
\[ \frac{1}{|\bP^{34}|} \, \sum_{i=1}^{11} N_i \, {\rm wt}(\omega_i) \,= \,|G(3,7)|  \left( 1- \frac{ |\bP^{33}|}{|\bP^{34}|} \right)\]
which has the interpretation that  the average weight of a $C(3,7)$ codeword  equals $|G(3,7)|$ times the fraction of points of $\bP^{34}$ not lying  on a fixed hyperplane. We verified this identity on a computer algebra system by evaluating the left hand side of
\eqref{eq:MacWilliams} using the weights from \eqref{eq:weights} and the $N_i$'s from \eqref{eq:N_i}.

\section{Concluding remarks: Spectrum of $C(3,m), \, m>7$}
Theorem \ref{weight theorem} allows us to calculate the weight of a non-degenerate codeword $\omega$ of $C(3,m)$ in terms of the cardinalities of its weight varieties $X_1(\omega), \cdots, X_{ \lfloor \frac{m-1}{2} \rfloor}(\omega)$. 
Proposition \ref{prop:degenerate_weight}  reduces the calculation of weights of degenerate codewords of $C(3,m)$ to that of non-degenerate codewords of $C(3,\tilde m)$ for $\tilde m <m$. The image of the function $\omega \mapsto $ wt$(\omega)$ from non-zero codewords to positive integers, and the number of pre-images of each integer in its image, is the spectrum of $C(3,m)$.
Since the number of non-zero codewords is $q^{\binom{m}{3}}-1$ is large, it is not feasible to evaluate the weight function for all codewords. The method proposed here for $m \leq 7$ is to use the fact that projectively or linearly equivalent codewords have the same 
weights, to evaluate the weights only on the projective or linear equivalence classes.
 Let $\nu: = \binom{m}{3} - m^2$. We note that 
$\nu < 0$ iff $m \leq 8$. Let $m >8$ and let $\gamma(m,q)$ denote the number of linear equivalence classes of $C(3,m)$ codewords. If $\omega_1, \cdots, \omega_{\gamma(m,q)}$ are representatives of these equivalence classes then:
\[ \sum_{i=1}^{\gamma(m,q)} \frac{|GL(m,\bF)|}{|{\rm Aut}(\omega_i)|} = q^{\binom{m}{3}}-1 \]
Since $|GL(m,\bF)|/|$Aut$(\omega_i)| \leq |GL(m,\bF)| $, and $|GL(m,\bF)| = q^{m^2} + O(q^{m^2-1})$ we get
\[ \gamma(m,q) \geq  q^\nu +O(q^{\nu-1}) \quad \mbox{for} \; m>8 \]
%  and therefore \[  \gamma(m,q) \in \Omega(q^{\nu})\]
%The number of projective equivalence classes is thus in $\Omega(q^{\nu-1})$.
%Due to this combinatorial explosion for $m>8$, the method of classifying the weights of projective classes of codewords is not effective.
The number of projective equivalence classes is thus greater than $q^{\nu-1} +O(q^{\nu-2})$.
Although the number of distinct weights is in general less than the number of projective classes, we believe that the former will still be bounded  below by polynomial function of $q$ for any fixed $m>8$.  \\

The problem of calculating the spectrum of the code $C(3,8)$ on the other hand is much more tractable. By Proposition \ref{prop:degenerate_weight} we need determine only the weights of non-degenerate $3$-forms.
%can be completely determined, since the number of linear equivalence classes of codewords for $q=p^j$ is independent of $j$. 
Noui \cite{Noui1} has shown that there are $13$ linear equivalence classes of non-degenerate $3$-forms over $\bar \bF^8$ where $\bar \bF$ is an algebraic closure of $\bF$.  Let $\omega_{8,1}, \cdots, \omega_{8,13}$ (in the notation of \cite{Noui1},\cite{Noui2}) be 
representative $3$-forms for these classes. Distinct  linear equivalence classes over $\bF$ may turn out to be linearly equivalent over $\bar \bF$. Following the method used by Cohen and Helminck \cite{Cohen} for $m=7$, once the groups
Aut$(\omega_{8,j}) \subset GL(8,\bar \bF)$ are known, the methods of Galois cohomology can be used to determine the classes
$\omega_{8,j}$ which split into multiple classes when going from $\bar \bF$ to $\bF$. This program is partially carried out by 
Noui and Midoune \cite{Noui2} (Corollary 2) for the first $6$ forms $\omega_{8,1}, \cdots, \omega_{8,6}$. Under the restriction char$(\bF) \neq 2,3$, they show explicitly that these $6$ classes over $\bar \bF$ yield $9$ classes over $\bF$. 
As future work one can complete this program, and use it to fully determine the spectrum of $C(3,8)$. 
%\section*{Acknowledgment}

\bibliographystyle{IEEEtran}
\bibliography{refs}{}

\begin{thebibliography}{1}
\providecommand{\url}[1]{#1}
\csname url@samestyle\endcsname
\providecommand{\newblock}{\relax}
\providecommand{\bibinfo}[2]{#2}
\providecommand{\BIBentrySTDinterwordspacing}{\spaceskip=0pt\relax}
\providecommand{\BIBentryALTinterwordstretchfactor}{4}
\providecommand{\BIBentryALTinterwordspacing}{\spaceskip=\fontdimen2\font plus
\BIBentryALTinterwordstretchfactor\fontdimen3\font minus
  \fontdimen4\font\relax}
\providecommand{\BIBforeignlanguage}[2]{{%
\expandafter\ifx\csname l@#1\endcsname\relax
\typeout{** WARNING: IEEEtran.bst: No hyphenation pattern has been}%
\typeout{** loaded for the language `#1'. Using the pattern for}%
\typeout{** the default language instead.}%
\else
\language=\csname l@#1\endcsname
\fi
#2}}
\providecommand{\BIBdecl}{\relax}
\BIBdecl

\bibitem{Nogin2}
D.~Y. Nogin, ``The spectrum of codes associated with the {G}rassmannian variety
  {$G(3,6)$},'' \emph{Problems of Information Transmission}, vol.~33, no.~2,
  pp. 114--123, 1997.

\bibitem{Cohen}
A.~M. Cohen and A.~G. Helminck, ``Trilinear alternating forms on a vector space
  of dimension {$7$},'' \emph{Comm. Algebra}, vol.~16, no.~1, pp. 1--25, 1988.

\bibitem{Ryan}
C.~Ryan, ``An application of {G}rassmannian varieties to coding theory,''
  \emph{Congr. Numer.}, vol.~57, pp. 257--271, 1987, sixteenth Manitoba
  conference on numerical mathematics and computing (Winnipeg, Man., 1986).

\bibitem{Nogin}
D.~Y. Nogin, ``Codes associated to {G}rassmannians,'' in \emph{Arithmetic,
  geometry and coding theory ({L}uminy, 1993)}.\hskip 1em plus 0.5em minus
  0.4em\relax Berlin: de Gruyter, 1996, pp. 145--154.

\bibitem{Tsfasman}
M.~Tsfasman, S.~Vl{\u{a}}du{\c{t}}, and D.~Nogin, \emph{Algebraic geometric
  codes: basic notions}, ser. Mathematical Surveys and Monographs.\hskip 1em
  plus 0.5em minus 0.4em\relax Providence, RI: American Mathematical Society,
  2007, vol. 139.

\bibitem{Lang}
S.~Lang, \emph{Algebra}, 3rd~ed., ser. Graduate Texts in Math.\hskip 1em plus
  0.5em minus 0.4em\relax New York: Springer-Verlag, 2002, vol. 211.

\bibitem{Sloane}
F.~J. MacWilliams and N.~J.~A. Sloane, \emph{The theory of error-correcting
  codes. {I}}.\hskip 1em plus 0.5em minus 0.4em\relax Amsterdam: North-Holland
  Publishing Co., 1977, north-Holland Mathematical Library, Vol. 16.

\bibitem{Noui1}
L.~Noui, ``Transvecteur de rang 8 sur un corps alg\'ebriquement clos,''
  \emph{C. R. Acad. Sci. Paris S\'er. I Math.}, vol. 324, no.~6, pp. 611--614,
  1997.

\bibitem{Noui2}
L.~Noui and N.~Midoune, ``{$K$}-forms of 2-step splitting trivectors,''
  \emph{Int. J. Algebra}, vol.~2, no. 5-8, pp. 369--382, 2008.

\end{thebibliography}
\nocite{}

\end{document}